\documentclass[twocolumn,rmp,amsmath,amssymb,showpacs]{revtex4}
\usepackage{graphicx}% Include figure files
\usepackage{dcolumn}% Align table columns on decimal point
\usepackage{bm}% bold math

\begin{document}

 \title{In-situ determination of the energy dependence of the high frequency mobility in polymers}

  \author{I. N. Hulea}
  \altaffiliation{present address: Kavli Instutute of Nanoscience, Delft University of Technology, The Netherlands; e-mail: I.Hulea@tnw.tudelft.nl}
  \affiliation{Kamerlingh Onnes Laboratory, Leiden University, P.O. Box 9504, 2300 RA Leiden, The Netherlands, and
  The Dutch Polymer Institute, P.O. Box 902, 5600 AX Eindhoven, The Netherlands}

  \author{A. V. Pronin}
  \altaffiliation{also at Institute of General Physics, Russian Academy of Sciencies, 119991 Moscow, Russia}

  \author{H. B. Brom}
  \affiliation{Kamerlingh Onnes Laboratory, Leiden University, P.O. Box 9504, 2300 RA Leiden, The Netherlands}

 \date{published online in Applied Physics Letters {\bf 86} on June 14}

\begin{abstract}
The high-frequency mobility ($\mu_{\rm hf}$) in disordered systems
is governed by transport properties on mesoscopic length scales,
which makes it a sensitive probe for the amount of local order.
Here we present a method to measure the energy dependence of
$\mu_{\rm hf}$ by combining an electrochemically gated transistor
with in-situ quasi-optical measurements in the sub-terahertz
domain. We apply this method to
poly([2-methoxy-5-(3',7'-dimethylocyloxy)]-$p$-phenylene vinylene)
(OC$_1$C$_{10}$-PPV) and find a mobility at least as high as 0.1
cm$^{2}$V$^{-1}$s$^{-1}$.
\end{abstract}


\maketitle

In many materials of practical interest, like doped or amorphous
semiconductors, metal-insulator composites, and doped conjugated
polymers, disorder introduces charge localization, and hence
charge transport is due to hopping of the carriers instead of band
conduction. Because of the disorder the mobility is often low. For
example, in polymeric light emitting diodes the dc hole mobility
($\mu_{\rm dc}$) for the often used conjugated polymer
poly([2-methoxy-5-(3',7'-dimethylocyloxy)]-$p$-phenylene vinylene)
(OC$_1$C$_{10}$-PPV) is typically of the order of $10^{-11}\rm\
m^2/Vs$ at room temperature.
\cite{Meyer95,Kryukov92,Gailberger91,Blom96,Hulea03} Because
$\mu_{\rm dc}$ in a disordered system is limited by the weakest
link in the conduction path, it usually can not provide detailed
insight in the transport processes involved on mesoscopic
lengthscales. As previously demonstrated for metal-insulator
composites, chemically doped polymers and polymers in solution,
measurements of $\mu$ as function of frequency are more
informative for the mesoscopic structure and the conduction
process.\cite{Adriaanse97,Hoofman98,Martens001} For example, a few
years ago Hoofman {\it et al.} applied their time-resolved
microwave conductivity technique to investigate the charge
mobility in a PPV solution.\cite{Hoofman98} In this method a high
energy electron beam (3 MeV) creates doping ions, and the
differences in transmission for 30 GHz radiation is translated to
the mobility, which was found to be about ${\rm 2 \cdot 10^{-5}\,
m^{2}/ Vs}$. Martens {\it et al.} used voltage-modulated
millimeter-wave spectroscopy to determine the carrier mobility in
solid state devices of OC$_1$C$_{10}$-PPV at frequencies between
10 and 200 GHz.\cite{Martens001} The carriers were introduced from
contacts in the device by a large electric field of around hundred
${\rm V/\mu m}$ and the high frequency mobility $\mu_{\rm hf}$,
deduced from the attenuation of the sub-THz radiation through the
device, was shown to be as high as ${\rm 10^{-4}\, m^2/Vs}$. The
dependence of ${\rm \mu_{hf}}$ on the applied voltage over the
contacts was linked to the occupation of the levels in the (almost
Gaussian) density of states (DOS) \cite{Bassler93}. This finding
was very recently supported by DOS data obtained with an
electrochemically gated transistor (EGT).\cite{Roest02,Hulea04}
The method was proven to be reliable for PPV and the mobility
could be followed over a relatively large energy scan of more than
an eV.\cite{Hulea04,Tanase03}

In this letter we describe a method to measure the high-frequency
mobility, which combines the process of electrochemical doping
using an EGT with in-situ quasi-optical measurements in the
sub-terahertz domain.\cite{Roest02,Hulea04} Knowing the amount of
doping and the precise energy calibration are great advantages
over the time-resolved microwave conductivity or voltage-modulated
millimeter-wave spectroscopy methods. Using OC$_1$C$_{10}$-PPV as
an illustration, we determine the dependence of the high-frequency
mobility on the electrochemical potential.

\begin{figure}[htb]
\begin{center}
 \includegraphics[height=6cm]{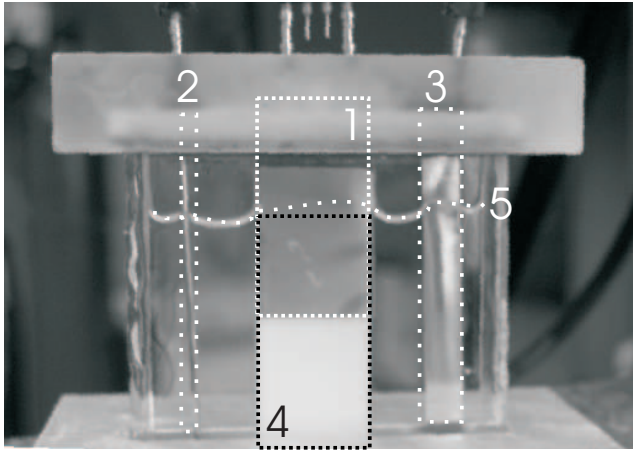}
\end{center}
\noindent{\caption{Front view of the electrochemical cell used.
The regions indicated by dotted lines are: 1 - sample (PPV on a
glass substrate), 2 - Ag reference, 3 - Pt counter electrode, 4 -
Teflon spacer positioned behind the sample, and 5 - level of the
electrolyte in the cell. The cell dimensions are about 5 cm
$\times$ 5 cm. A cap with a rubber O-ring seals the cell. Four
feed-through connections in the cap enable electrical contact with
the reference and counter electrodes and the source and drain
contacts on the sample. The direction in which radiation
propagates is perpendicular to the plane of the figure.}
\label{cell}}
\end{figure}

A major problem to overcome when working with ionic solutions is
that ions in the electrolyte strongly absorb at sub-THz
frequencies. In earlier studies with similar electrolytes and the
polymer dissolved in the solution, the absorption by the
electrolyte completely dominated and prevented investigation of
the properties of the polymer films. \cite{Reedijk99} To overcome
this difficulty, the electrochemical cell was designed to minimize
the path of the radiation in the solution. In our experiments the
radiation frequencies are around 190 GHz. The cell has been made
out of quartz and had a free space of 2.5 mm in the direction the
radiation propagates. Still, when filled with electrolyte it
absorbs more than 100 dB, which makes the measurements impossible.
To minimize the absorption of the electrolyte, a 1 mm teflon or
quartz spacer (for both materials the absorption is very low) was
mounted behind the 0.5 mm thick sample (glass plus film), see
Fig.~\ref{cell}. By this procedure the transmission loss could be
kept to an acceptable level. For the undoped sample at 190 GHz the
suppression of the transmission was found to be of the order of 50
dB.

\begin{figure}[htb]
\begin{center}
 \includegraphics[width=8cm]{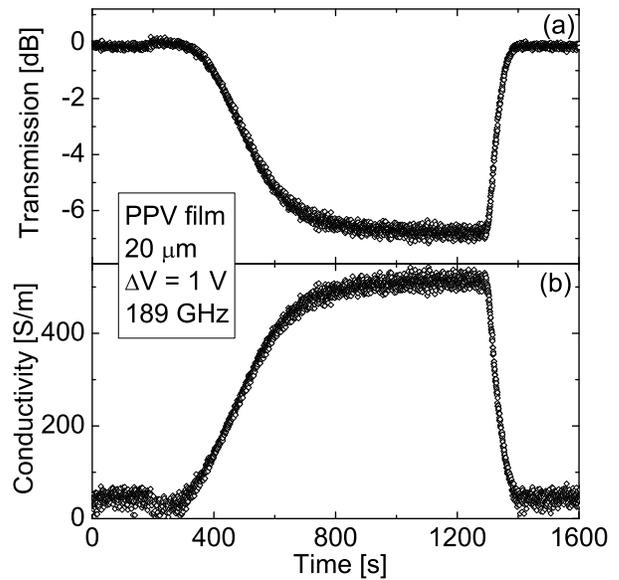}
\end{center}
\noindent{\caption{Upper panel: relative transmission of sub-THz
radiation measured as a function of time for a 20 $\mu$m thick PPV
sample. At t=0 the measurement is started with no potential
difference between the ITO electrode and the quasi-reference Ag
electrode, and the transmission here is chosen as 0 dB. After
about 200 s the potential on the ITO is raised suddenly while the
transmission is monitored continuously. Once the steady
transmission value is reached the difference in potential is
brought back to 0 V and the original value in transmission
(corresponding to the undoped sample) is recovered. Bottom panel:
conductivity calculated from the measured transmission via
Eqn.~\ref{cond}.} \label{fig2}}
\end{figure}

In the EGT cell a counter electrode (Pt foil) and a reference
electrode (Ag wire) are required, see Fig.~\ref{cell}, for
calibrated changes in the electrochemical potential. Because PPV
is air-sensitive, the cell has to be air-tight. Casting of the
samples on ITO covered glass and mounting in the cell were done
under nitrogen atmosphere.

The measuring procedure is straightforward: the electrochemical
cell is placed between the source and detection horns of the
sub-THz setup and by creating a potential difference ($\Delta V$)
between the working electrode (ITO) and the reference electrode
(Ag) doping is achieved. Results of a typical measurement, where
the transmission is monitored as function of time, are shown in
Fig.~\ref{fig2}a. To get a measurable difference in transmission
between the undoped and doped OC$_1$C$_{10}$-PPV sample, the
thickness of the PPV layer had to be more than 10 micron.
Initially $\Delta V$ is zero, that means no doping. After about
200 seconds $\Delta V$ is changed to a value of 1 V. Once the
steady-state value in transmission is achieved ($ \sim 10^3$
seconds after the potential change) $\Delta V$ is set to 0 V,
corresponding again to an undoped sample. While doping, we observe
that the transmission decreases by 6-7 dB and on switching back to
a non-doped state (0 V) the original transmission value is
recovered. Repeating this procedure several times we obtained the
transmission changes corresponding to different values of the
applied potential on the working electrode. From the measured
transmission the conductivity $\sigma$ can be calculated via
\begin{equation}\label{cond}
  |T|=\frac{A}{1+[(d/2c\varepsilon_0)\sigma(\omega)]^2} ,
\end{equation}
with $A$ a constant that takes into account the transmission of
the substrate, electrolyte, cell's walls, etc., $d$ is the polymer
film thickness, $c$ the speed of light, $\varepsilon_0$ the
permittivity of free space and $\sigma(\omega)$ the film
conductivity. Eqn.~\ref{cond} is only valid, if the optical
thickness of the film is less than the wavelength of the probing
radiation, which is satisfied in our case for films with thickness
less than 25 $\mu$m. The calculated conductivity as function of
time is shown in Fig.~\ref{fig2}b.

\begin{figure}[t]
\begin{center}
 \includegraphics[width=\columnwidth]{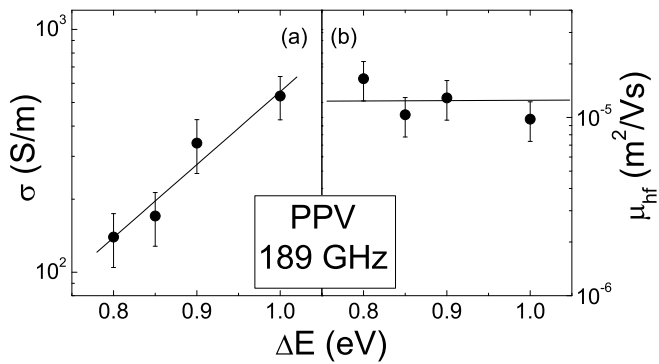}
\end{center}
\noindent{\caption{(a) The conductivity vs. energy. The energy is
relative to the Ag pseudo-reference, which is 4.47 V below the
vacuum level (the level of 1 eV is at -5.47 eV below the vacuum
level). (b) The mobility determined from the conductivity data of
the left panel, assuming a homogeneous loading of the film and all
doped charges to participate. The data are for PPV at 189 GHz;
lines are guides to the eye.} \label{fig3}}
\end{figure}

The measured conductivities for different potentials, see
Fig.~\ref{fig3}a, compare well with those of organic crystals or
amorphous silicon, with typical values of the order of 300 S/m.
From the conductivity, the high frequency mobility can be
extracted if we know the number of participating carriers. This
number depends on the density of states and the electrochemical
potential. In dc-transport, which is an interchain and intrachain
process, due to the Fermi-Dirac statistics especially transitions
in a band of $k_BT$ around the chemical potential contribute to
the conductivity.\cite{Hulea04} Close to a THz, most charges
oscillate within a segment of the chain, and the number of
contributing charges will be higher (for polypyrrole only in the
mid-infrared response all charges participate \cite{Romijn03}).
The number of all doped charges is obtained by integrating the
density of states known from previous experiments on thin
PPV-devices,\cite{Hulea04} and will be used here for estimating
(the lower limit) of $\mu_{\rm hf}$. When the electrochemical
potential with respect to the Ag reference equals 0.8 eV, the
mobility is slightly higher than $10^{-5}$ m$^2$/Vs. A similar
value is deduced at half band filling (at an electrochemical
potential of 1 eV \cite{Hulea04}). These results reproduce the
findings of Hoofman {\it et al.}  within a factor
two,\cite{Hoofman98} which shows that at low doping the frequency
dependence of $\mu_{\rm hf}$ between 30 GHz and 200 GHz is
negligible. Qualitatively these findings confirm the ultra-fast
oscillatory motion of the carriers in the chain segment. As a
comparison, the mobilities obtained in the best organic crystals
are around $10^{-3}$ m$^2$/Vs, \cite{Stassen04,Sundar04} i.e. two
orders of magnitude higher than reported here for low doping
levels.

In samples of 0.2 $\mu$m the typical time constant for the
transmission change was of the order of 0.2 s or at least a factor
of $10^3$ shorter than the 200 - 600 s for the 20 $\mu$m film, see
Fig.~\ref{fig2}. Differences of that order of magnitude are
expected in diffusion processes, where time goes with the mean
distance squared; a real computation would have to account for the
opening of the diffusion channels (swelling) and the history of
the experiment.

In conclusion, we have developed a method to measure the high
frequency mobility based on an electrochemically gated transistor.
The advantages over the existing methods are its ability to deal
with solid state films, higher doping levels than the ones
obtained in diode configurations, and the precise energy and
concentration calibration. The results show that the structure at
short length scales supports a mobility which comes close to those
of organic crystals.

It is a pleasure to acknowledge Arjan Houtepen, Dani\"el
Vanmaekelbergh and John Kelly (Utrecht University) for their
advise and support. This work forms part of the research program
of the Dutch Polymer Institute DPI, project nr. 274.

\end{document}